\documentclass[namedreferences]{kluwer}
\usepackage{epsfig}
\newcommand{\magpt}[2]{\mbox{$\rm #1\hspace{-0.25em}\stackrel{m}{.}
      \hspace{-1.0mm}#2$}}                             
\newcommand{\magn}[1]{\mbox{$\rm #1\hspace{-0.05em}^m$}}

\newcommand\teff{$ {\rm T_{eff}}$}

\newcommand\loghe{${\rm \log{\frac{n_{He}}{n_{H}}}}$}

\def\gtrsim{\mathrel{\hbox{\rlap{\hbox{\lower4pt\hbox{$\sim$}}}\hbox{$>$}}}}
\def\lesssim{\mathrel{\hbox{\rlap{\hbox{\lower4pt\hbox{$\sim$}}}\hbox{$<$}}}}

\newcommand{\Msolar}{\mbox{\,$\rm M_{\odot}$}}        
\begin{document}
\begin{article}
\begin{opening}
\title{Helium-rich EHB Stars in Globular Clusters\thanks{Based on
observations at ESO (66.D-0199, 68.D-0248)}}
\author{S. \surname{Moehler}\email{moehler@astrophysik.uni-kiel.de}}
\institute{Institut f\"ur Theoretische Physik und Astrophysik der Universit\"at
Kiel, Abteilung Astrophysik, 24098 Kiel, Germany}
\author{A.V. \surname{Sweigart}} \author{W.B. \surname{Landsman}}
\institute{NASA\,Goddard Space Flight
Center, Code 681, Greenbelt, MD 20771, USA}
\author{S. \surname{Dreizler}}
\institute{Astronomisches Institut der
Universit\"at T\"ubingen, Sand 1, D-72076 T\"ubingen, Germany}
\runningtitle{Helium-rich EHB Stars in Globular Clusters}
\runningauthor{Moehler et al.}
\begin{abstract} 
Recent UV observations of the most massive Galactic globular clusters
show a significant
population of hot stars below the zero-age HB (``blue hook'' stars),
which cannot be explained by canonical stellar evolution.  Stars which
suffer unusually large mass loss on the red giant branch and thus
experience the helium-core flash while descending the white dwarf
cooling curve could populate this region.  They should show higher
temperatures than the hottest canonical HB stars and their atmospheres
should be helium-rich and probably C/N-rich.
We have obtained spectra of blue hook stars in $\omega$ Cen
and NGC 2808 to test this possibility.  Our analysis shows
that the blue hook stars in these clusters reach effective
temperatures well beyond the hot end of the canonical EHB and
have higher helium abundances than canonical EHB stars.  These
results support the hypothesis that the blue hook stars arise
from stars which ignite helium on the white dwarf cooling curve.
\end{abstract}




\end{opening}
\section{Introduction}
Low-mass stars burning helium in a core of about 0.5~\Msolar\ and
hydrogen in a shell populate a roughly horizontal region in the
colour-magnitude diagrams of globular clusters, which has earned them
the name ``horizontal branch" ({\bf HB}) stars.  The Galactic globular
clusters show a great variety in horizontal branch morphology, i.e. in
the temperature distribution of their HB stars.  The temperature of an
HB star depends -- at a given metallicity -- on the mass of its
envelope, with the hottest or extreme HB (EHB) stars (\teff\ $>$
20,000~K) having extremely thin ($\lesssim$0.01\Msolar) envelopes.
For stars hotter than about 10,000~K the increase in the bolometric
correction with increasing temperature turns the blue HB into a
vertical blue tail in optical colour-magnitude diagrams, with the
faintest blue tail stars being the hottest and least massive. EHB
stars are in general fainter than $M_V \approx$ \magn{+4} and thus
most of the stars along the blue tails are classical hot horizontal
branch stars.

Observations of some of the most massive Galactic globular clusters in
the far- and near-UV ($\omega$ Cen, Whitney et al. \citeyear{whro98},
D'Cruz et al. \citeyear{dcoc00}; NGC~2808, Brown et
al. \citeyear{brsw01}; NGC~6388, NGC~6441, Busso et
al. \citeyear{bupi03}) show a group of stars forming a hook-like
feature up to \magpt{0}{7} below the hot end of the zero-age HB
(ZAHB). Within the framework of canonical HB theory there is no way to
populate this region of the UV colour-magnitude diagram
 without requiring an implausibly
large decrease in the helium-core mass.

Optical colour-magnitude diagrams of these globular clusters show that the
blue hook stars lie at the very faint and hot end of the blue
tails ($M_V \approx$ \magpt{4}{5}--\magpt{5}{5}). That in itself
would not be a problem, but the spectroscopic analyses of Moehler et
al. (\citeyear{mosw00}) show that the blue tail stars in NGC~6752
(which are all brighter than $M_V \approx$ \magpt{4}{5}) already
populate the EHB to the hot end predicted by canonical HB models.
Thus canonical theory fails to explain both the faint UV luminosities
and expected high temperatures of the blue hook stars.

Stars which lose a large amount of mass on the red giant branch can
leave the red giant branch without experiencing the helium flash, and
move quickly to the (helium-core) white dwarf cooling curve.
Castellani \& Castellani (\citeyear{caca93}) were the first to suggest
that - for very high mass loss on the RGB - the helium flash can occur
at high effective temperatures after a star has left the RGB (the
so-called ``hot flashers''), either during the evolution to the top of
the white dwarf cooling curve (``early hot flashers'') or while
descending the white dwarf cooling curve (``late hot flashers'').
Indeed, D'Cruz et al. (\citeyear{dcdo96}, \citeyear{dcoc00}) proposed
that the blue hook stars could be the progeny of such hot flashers,
but unfortunately the D'Cruz et al. models were, at most, only
$\approx$\magpt{0}{1} fainter than the canonical ZAHB, much less than
required by the observations.  More recently, Brown et
al. (\citeyear{brsw01}) have explored the evolution of the hot
flashers through the helium flash to the EHB in more detail,
especially with regard to the timing of the flash. Their models show
that a late hot helium flash on the white dwarf cooling curve will
induce substantial mixing between the hydrogen envelope and the helium
core, leading to helium-rich EHB stars that are much hotter than
canonical ones. They suggest that this flash mixing may be the key for
understanding the evolutionary status of the blue hook stars. Such
mixing may also be responsible for producing the helium-rich, high
gravity field sdO stars (Lemke et al. \citeyear{lehe97}), whose origin
is otherwise obscure.  The flash mixing scenario predicts a helium
dominated atmospheric composition for the late hot flashers as well as a gap of
about 6000~K between the late hot flashers and 
canonical EHB stars (i.e. EHB stars which have
not experienced flash mixing and which therefore
have hydrogen-rich atmospheres).

\begin{figure}[!ht]
\centerline{\includegraphics[width=12cm]{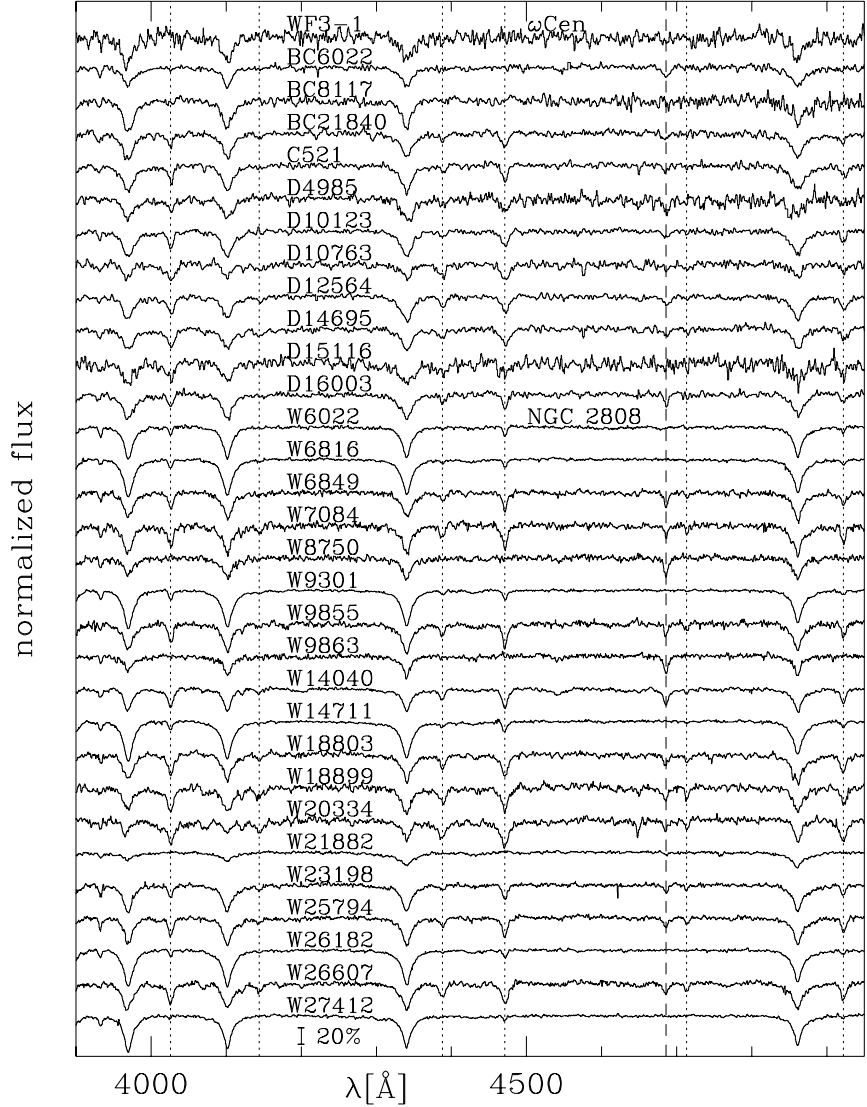}}
\caption[]{Our spectra of blue tail/blue hook stars in $\omega$~Cen
(WF3$-$1 was identified as a target in the HST photometry of D'Cruz et
al. \citeyear{dcoc00}, all other numbers refer to the photometry
from Kaluzny et al. \citeyear{kaku96},\citeyear{kaku97}) and NGC~2808
(numbers refering to the photometry of Walker 1999). The dotted lines
mark the position of the He~{\sc i} lines, the dashed line marks the
He~{\sc ii} line at 4686~\AA.  }\label{spectra}
\end{figure}

\section{Observations and Data Reduction}\label{sec-obs}

We obtained medium-resolution spectra ($R\approx$700) of 12 blue hook
candidates in $\omega$ Cen with 18.5 $< V <$ 19.2 at the NTT with EMMI
on February 22--25, 2001 and medium-resolution spectra ($R\approx$900) of
19 stars along the blue tail in NGC~2808 with \magn{19} $< V <$
\magn{21} at the VLT-UT1 (Antu) with FORS1 in service mode. The data
reduction for the $\omega$ Cen data is described in Moehler et
al. (\citeyear{mosw02}).  The data reduction for the NGC~2808 data
will be described in Moehler et al. (\citeyear{mosw03}).
The spectra, given in Fig.~\ref{spectra}, show a large variety of helium
line strengths. Part of this variation is due to variations in
effective temperature, which is evident from the varying strength of
the Balmer absorption lines. However, as we will see in the analysis,
the helium abundance also varies considerably.

\section{Analysis}

\begin{figure}[ht]
\centerline{\includegraphics[width=10cm]{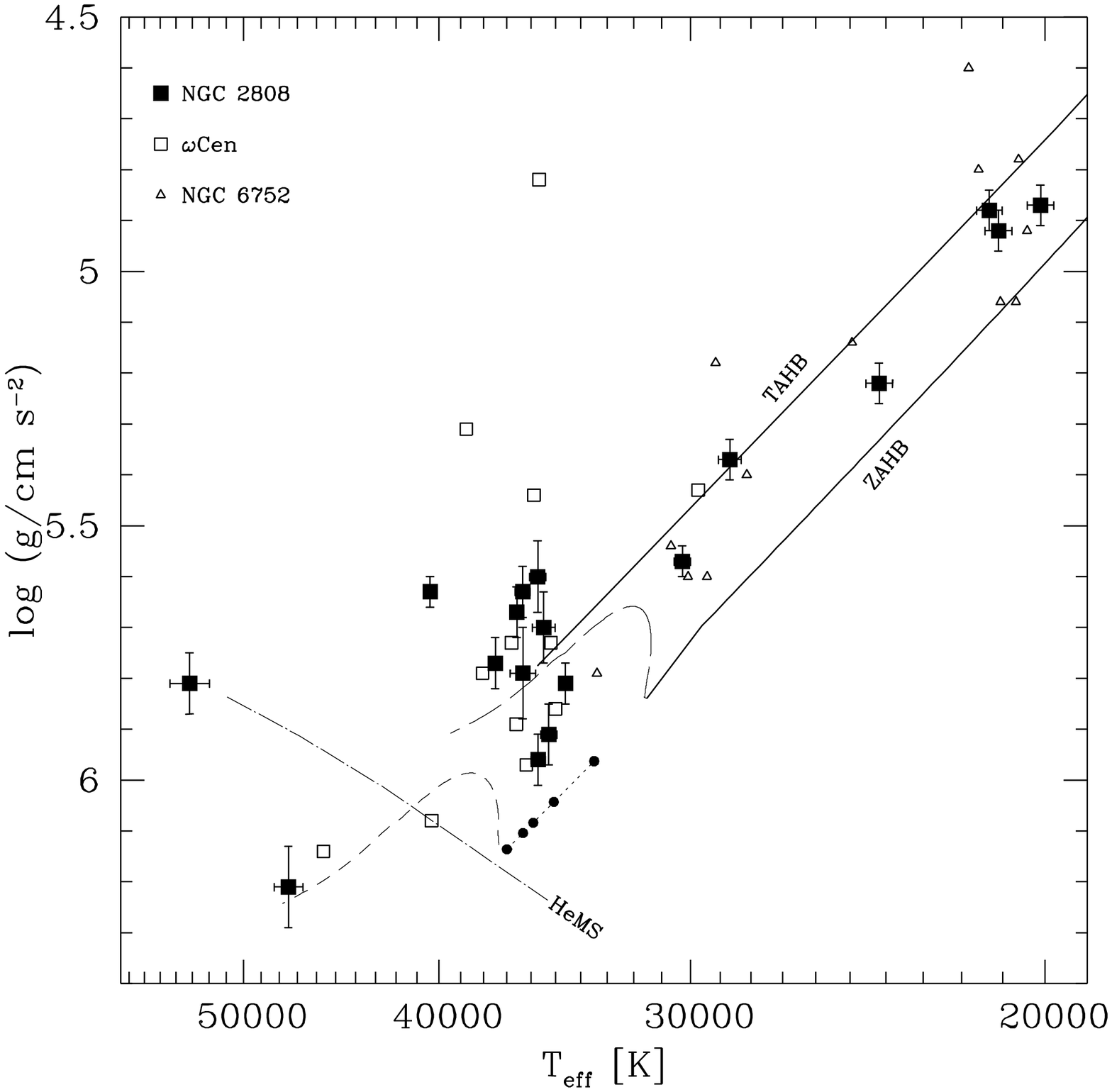}}
\caption[]{Atmospheric parameters 
derived from the spectra of stars along the blue tail/blue hook
in NGC~2808 (filled squares) and $\omega$ Cen (open squares)
compared to HB evolutionary tracks. Also
shown are blue tail stars from NGC~6752 (open triangles, Moehler et
al. \citeyear{mosw00}).  The tracks for an early hot
flasher (long-dashed line) and a late hot flasher (short-dashed
line) show the evolution of such stars from the zero-age HB (ZAHB)
towards helium exhaustion in the core (terminal-age HB = TAHB). The
solid lines mark the canonical HB locus for [M/H] = $-$1.5 from
Sweigart (\citeyear{swei97}). The dotted line connects the series of ZAHB
models computed by adding a hydrogen-rich layer to the surface of the
ZAHB model of the late hot flasher.  The filled circles mark --
with decreasing temperature -- hydrogen layer masses of $0, 10^{-7},
10^{-6}, 10^{-5}, 10^{-4}$\Msolar. The dash-dotted line marks the
helium main sequence of Paczynski \citeyear{pacz71})}\label{tg}
\end{figure}
To obtain effective temperatures, surface gravities and helium
abundances we fitted the spectra with theoretical model spectra. For 
hot stars (\teff $\gtrsim$33,000~K) we used non-LTE H-He model
atmospheres as described in Moehler et al. (\citeyear{mosw02}).
 For the cooler stars we used ATLAS9 line blanketed LTE
model atmospheres for solar metallicity (to account for effects of
radiative levitation), from which we calculated spectra with Lemke's
version\footnote{For a description see\\
http://a400.sternwarte.uni-erlangen.de/$\sim$ai26/linfit/linfor.html}
of the LINFOR program (developed originally by Holweger, Steffen, and
Steenbock at Kiel University). To establish the best fit we used the
routines developed by Bergeron et al.\ (\citeyear{besa92}) and Saffer et
al.\ (\citeyear{sabe94}), as modified by Napiwotzki et
al. (\citeyear{nagr99}), which employ a $\chi^2$ test. The $\sigma$
necessary for the calculation of $\chi^2$ is estimated from the noise
in the continuum regions of the spectra. The fit program normalizes
model spectra {\em and} observed spectra using the same points for the
continuum definition. During the analysis of the data presented here
we realized that helium-poor and helium-rich spectra required
different sets of continuum points. Especially the use of the
continuum points derived from hydrogen-rich spectra for the analysis
of helium-rich spectra can introduce large systematic errors,
esp. overestimates of the helium abundance due to continuum points in
the wings of strong helium lines and/or too narrow fitting windows for
helium lines. We therefore refined the definition of the continuum
points and re-analysed the data for the blue hook
stars in $\omega$ Cen presented in Moehler et al. (\citeyear{mosw02}).
The new results differ from the old ones usually within the mutual
error bars, except for the two most helium-rich stars, as expected.

\begin{figure}[ht]
\centerline{\includegraphics[width=10cm]{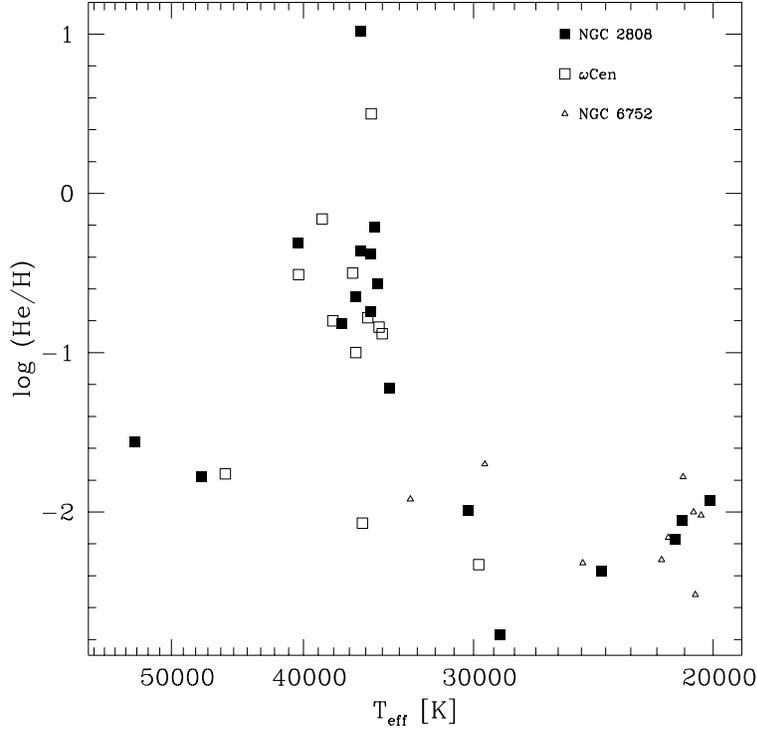}}
\caption[]{Helium abundances vs. effective temperature
for the stars along the blue tail/blue hook
in NGC~2808 (filled squares) and $\omega$ Cen (open squares). Also
shown are blue tail stars from NGC~6752 (open triangles, Moehler et
al. \citeyear{mosw00}). Solar helium abundance corresponds to \loghe
$\approx -1$. 
}\label{the}
\end{figure}

We used the Balmer lines H$_\beta$ to H$_{8}$ (excluding H$_\epsilon$
to avoid the Ca~{\sc ii}~H line), the He~{\sc i} lines
$\lambda\lambda$ 4026~\AA, 4388~\AA, 4471~\AA, 4921~\AA\, and the
He~{\sc ii} lines $\lambda\lambda$ 4542~\AA, 4686~\AA\ for the
helium-poor stars. For the helium-rich stars we also included the
He{\sc i} lines $\lambda\lambda$ 4713~\AA, 5015~\AA\ and 5044~\AA\ in
the fit. The results are plotted in Figs.~\ref{tg} and \ref{the}.

\section{Discussion}
Our analysis of the blue hook stars in NGC~2808 and $\omega$ Cen
shows that these stars do indeed reach effective temperatures of more
than 35,000 K (cf. Fig.~\ref{tg}), well beyond the hot end of the
canonical EHB. In addition, most of the hot stars show at least solar
helium abundances with the helium abundance increasing with effective
temperature (cf. Fig.~\ref{the}), in contrast to canonical EHB stars
such as those studied in NGC~6752 by Moehler et
al. (\citeyear{mosw00}).  We now discuss both of these results in more
detail.

Contrary to the predictions of Brown et al. (\citeyear{brsw01}) and
Cassisi et al. (\citeyear{casc03}), the atmospheres of the blue hook stars
still show some hydrogen. 
This result has been discussed by Cassisi et al. (\citeyear{casc03}), who find
that the mixing efficiency must be reduced by a factor of about 20,000
in order to reproduce even the highest helium abundances observed by Moehler
et al. (\citeyear{mosw02}).
However, this reduction only applies if the observed helium abundances
reflect the actual helium abundances in the envelopes of the
blue hook stars. 

The remaining hydrogen could be possibly explained by the outward diffusion of
hydrogen into the atmospheres of the blue hook stars and the
gravitational settling of helium.  Such diffusive processes are
believed to be responsible for the low helium abundances of the sdB
stars and are estimated to operate on a time scale much shorter than
the HB lifetime. 
It is unclear, however, whether diffusion might be inhibited to some
extent by convection within the helium-enriched atmospheres of the
flash-mixed stars.  Groth et al. (\citeyear{grku85}) found that
atmospheric convection can exist in hot subdwarfs if the helium
abundance is sufficiently high.  The range in the hydrogen abundances
of the blue hook stars might therefore indicate that varying amounts
of hydrogen survive flash mixing or that the efficiency of diffusion
differs from star to star. In any case the high helium abundances
observed in some of the blue hook stars would be difficult to
understand if their atmospheres were not enriched in helium during the
helium flash. The increase in the mean atmospheric helium abundance
with increasing effective temperature is also consistent with flash
mixing.

The presence of a hydrogen-rich surface layer would shift the
evolutionary track for the late hot flasher in Fig.~3
towards cooler temperatures.  This evolutionary track, taken from the
blue hook sequences of Brown et al. (\citeyear{brsw01}), has a
helium/carbon-rich envelope with no hydrogen.  In order to estimate
the size of this temperature shift, we computed a series of ZAHB
models in which hydrogen-rich layers with masses of 10$^{-7}$,
10$^{-6}$, 10$^{-5}$ and 10$^{-4}$~M$_\odot$ were added to the ZAHB
model from the late hot flasher in Fig.~\ref{tg}.  A hydrogen
layer of 10$^{-4}$~M$_\odot$ corresponds to the case in which
$\approx$10\% of the envelope hydrogen survives flash mixing and
in which all of this hydrogen then diffuses to the surface.
As expected, the ZAHB location of the late hot flasher in
Fig.~3 shifts redward as the mass of the hydrogen layer
increases and we see that the addition of a hydrogen layer of $<$
10$^{-4}$~M$_\odot$ would actually improve the agreement between the
predicted and observed temperatures of the blue hook stars while at
the same time preserving the temperature gap between these stars and
the canonical EHB stars.
While the HB track for the early hot flasher in Fig.~\ref{tg} passes
through the temperature gap, stars evolve very fast along this part of
the track and the probability to find evolved stars in that reagion is
very low.

We therefore conclude that
the high temperatures and high helium abundances reported here for the
blue hook stars in NGC~2808 and $\omega$ Cen provide general support
for the flash-mixing hypothesis of Brown et
al. (\citeyear{brsw01}). 

\begin{acknowledgements} 
We want to thank the staff at Paranal for their efforts in performing
the observations and R. Napiwotzki for his model atmospheres. 
We thank N. Hammer for his help in
calculating the model atmosphere grid.
\end{acknowledgements}

\end{article}%
\end{document}